# Open and reusable deep learning for pathology with WSInfer and QuPath

Jakub R. Kaczmarzyk[a], Alan O'Callaghan[b], Fiona Inglis[b], Tahsin Kurc[a], Rajarsi Gupta[a], Erich Bremer[a], Peter Bankhead[b,c,*], Joel H. Saltz[a,*]

[a] Department of Biomedical Informatics, Stony Brook University, Stony Brook, NY, USA
[b] Centre for Genomic & Experimental Medicine, Institute of Genetics and Cancer, The University of Edinburgh, Edinburgh, UK
[c] Edinburgh Pathology and CRUK Scotland Centre, Institute of Genetics and Cancer, The University of Edinburgh, Edinburgh, UK
* Peter Bankhead and Joel H. Saltz should be regarded as co-senior and co-corresponding authors.

## Abstract

The field of digital pathology has seen a proliferation of deep learning models in recent years. Despite substantial progress, it remains rare for other researchers and pathologists to be able to access models published in the literature and apply them to their own images. This is due to difficulties in both sharing and running models. To address these concerns, we introduce WSInfer: a new, open-source software ecosystem designed to make deep learning for pathology more streamlined and accessible. WSInfer comprises three main elements: 1) a Python package and command line tool to efficiently apply patch-based deep learning inference to whole slide images; 2) a QuPath extension that provides an alternative inference engine through user-friendly and interactive software, and 3) a model zoo, which enables pathology models and metadata to be easily shared in a standardized form. Together, these contributions aim to encourage wider reuse, exploration, and interrogation of deep learning models for research purposes, by putting them into the hands of pathologists and eliminating a need for coding experience when accessed through QuPath. The WSInfer source code is hosted on GitHub and documentation is available at https://wsinfer.readthedocs.io.

## Introduction

Pathology is the bedrock of cancer diagnosis and traditionally relies on the examination of physical slides containing human tissue specimens using high-power microscopy. In recent years, the field has been moving towards digital pathology, whereby glass slides are scanned as high-resolution images, known as whole slide images (WSIs). Each individual WSI is typically very large, often over 40 gigabytes uncompressed. The widespread adoption of digital pathology therefore poses considerable challenges for data storage and visualization, but also unlocks the potential to apply computational methods for diagnostics and prognostics.

It is difficult to overstate the transformative effect deep learning has had on digital pathology research. Many studies have suggested the potential for deep learning-based AI methods to revolutionize different aspects of pathology practice, such as by reducing the pathologist's workload or by augmenting visual assessment with the ability to identify subtle, sub-visual features of clinical importance (1–3). However, the multitude of algorithms published in the literature belies a dearth of implementations that are actually usable within the research community. In most cases, it is simply not possible for other research groups to validate the use of published methods on their own images and cohorts. One reason for this is that required data is not available: a recent survey of 161 peer-reviewed studies using deep learning for pathology found that while 1 in 4 shared code, only 1 in 8 shared trained model weights (4,5). Furthermore, in the minority of cases where code and models are available, they are typically not in a form amenable to pathologists without coding experience to use and explore. The result is that reported findings cannot properly be reproduced and interrogated by the wider community, and the key domain experts — pathologists — often find themselves to be particularly excluded. Tackling problems such as model generalization and overcoming batch effects urgently requires an increase in openness, replicability, and reusability.

In the present paper, we respond to the call to "make deep learning algorithms in computational pathology more reproducible and reusable" (4) by introducing WSInfer (Whole Slide Inference): a new collection of software tools designed to streamline the sharing and reuse of trained deep learning models in digital pathology (Figure 1).

We have focused on the generic task of patch classification, which is widely used across a broad range of pathology applications. Because WSIs are so big, they are typically broken into manageable patches to make analysis practicable. Trained patch-based deep neural networks are typically applied across a WSI to classify patches into different tissue components (e.g. tumor, stroma, fat, necrosis) or make predictions directly related to patient outcome. While relatively coarse-grained in comparison to an analysis based on segmenting individual structures, patch classification algorithms have advantages both in terms of computational efficiency and being a closer match for a pathologist's visual assessment — since this is often based upon evaluating patterns and textures, rather than discrete quantifiable entities. The output of patch classification is typically a spatial classification map, which can often be integrated across the WSI to create a single output representing a diagnosis, prediction, or 'score' for that slide.

## Description

WSInfer comprises three main components: (1) the WSInfer inference runtime, (2) the QuPath WSInfer extension, and (3) the WSInfer Model Zoo. Together these provide tools designed to meet the needs of a diverse range of users, including pathologists, computational researchers, and data scientists.

**Inference Runtime**

The WSInfer inference runtime deploys trained patch classification deep learning models on whole slide images and is available as a command line tool and Python package. The inference runtime requires three inputs from the user: a directory of whole slide images, a trained patch classification model, and a directory in which to write results. One may use a model from the Zoo or provide a local trained model along with a configuration JSON file that includes essential information for model use (i.e., size and physical spacing of patches, processing steps, names of output classes). The configuration file is validated against a schema to aid users in creating this file. If using a model from the Zoo, the model and configuration JSON file are downloaded automatically from the Hugging Face Hub. Each whole slide image undergoes a series of processing steps that were motivated by (6). First, patches are extracted from tissue regions at a uniform size and physical spacing, and each patch is processed as specified in the configuration JSON file (e.g., resized, normalized). An important optimization in this stage is the lazy loading of patches directly from the whole slide image. Compared to saving patches as image files, lazy loading requires less storage and performs fewer reads and writes to the filesystem. WSInfer offers a choice of slide reading backends between OpenSlide (7) and TiffSlide (8). Next, the patches are run through the forward pass of the deep learning model. Patches are loaded in parallel using the PyTorch DataLoader object. The runtime saves model outputs in comma-separated values (CSV) files with descriptive column names and GeoJSON files, a common format for spatial data. These output files can be used for downstream analyses or visualized using other software, including QuPath. The runtime can be installed with pip or as a Docker or Apptainer container.

We measured the running time of WSInfer in two environments: 1) a RedHat Linux environment with an enterprise-grade GPU (Quadro RTX 8000) and 2) a Windows Subsystem for Linux environment (Windows 11 and Debian 12) with a consumer GPU (RTX 2080 Ti). In both cases, we used the breast tumor classification model "breast-tumor-resnet34.tcga-brca" from the WSInfer Model Zoo (described below) and WSIs from The Cancer Genome Atlas. The model uses 350x350-pixel patches at 0.25 micrometers per pixel. In the RedHat Linux environment, analysis of 1,061 slides took 6 hours and 46 minutes, or ***23 seconds per WSI***. The distribution of the number of patches across WSIs was right-skewed (min=884, max=82,012, median=22,656, mean=23,492, std. dev.=13,922). In the second environment, we deployed the same model to 30 WSIs, a subset of the 1,061 used above. The running time was 14 minutes and 17 seconds total, or ***29 seconds per WSI***, and the distribution of patch counts was skewed similarly to the first example (min=6,575, max=52,323, median=23,502, mean=26,667, std. dev.=13,466).

**QuPath Extension**

QuPath is a popular open-source software platform for bioimage analysis (9). QuPath's support for visualizing, annotating, and analyzing whole slide images has led to the software being widely adopted within the digital pathology community: to date, it has been downloaded over 400,000 times and cited in over 2,400 studies. We therefore developed the QuPath WSInfer Extension as an alternative inference engine to make patch-based classification widely accessible within a familiar, intuitive, and interactive user interface.

The QuPath WSInfer Extension introduces patch-based deep learning support to QuPath the first time, building upon the software's existing features to provide an end-to-end analysis solution. Users are guided through the steps of selecting a deep learning model and one or more regions of interest for inference. The extension will then proceed to download the model if required, generate tile objects, and run inference (powered by Deep Java Library and PyTorch) at the appropriate resolution and patch size – appending the model output to the tiles. The user can then visualize the tile classifications and view interactive maps of predicted class probabilities. Furthermore, the tiles can be reused to run inference using additional models, making it possible to integrate information across models. Finally, because the user has access to all QuPath's other features (e.g. for tile merging, cell segmentation, data export), WSInfer can be integrated into sophisticated QuPath analysis pipelines, which are run either interactively or through automated scripts.

**Model Zoo**

We have curated a collection of trained pathology models for broad, unencumbered reuse and have hosted this Zoo on Hugging Face Hub. Each model repository contains a model card (10), pretrained weights in TorchScript format, and a configuration JSON file. The model card is a markdown file with human-readable metadata including the purpose of the model, its architecture, description of training data, how to apply it to new data, intended uses, and relevant citations. TorchScript is a serialization format that contains weights and a graph of the forward pass of the model, and it allows the use of the model without a Python dependency. The WSInfer QuPath extension, for instance, uses TorchScript models in a Java ecosystem. To add a model to the zoo, one creates a new model repository on Hugging Face Hub and uploads a model card, TorchScript file of the model, and configuration JSON file. One may optionally upload other files as well. Crucially, the user owns the model repository and can license and manage the contents independently. The registry of models in the zoo is maintained as a JSON file in a dedicated public repository on Hugging Face Hub. After publishing a model on Hugging Face Hub, one may submit a pull request to this repository adding the model location to the registry.

We have also developed a client utility to enhance interoperability of the zoo with other software. The client is available as a Python package or command-line tool and primarily lists and downloads models from the zoo. The client can also validate Model Zoo repositories and model configuration JSON files, functionalities we hope will ease the use of WSInfer.

## Discussion

WSInfer provides an open-source, cross-platform, and cross-language ecosystem to make deep learning methods uniquely accessible and intuitive for a wide range of digital pathology stakeholders. The core inference runtime is developed in Python, making it readily accessible for data scientists and deep learning specialists working in digital pathology — for whom Python is typically the programming language of choice. However, by also providing a Java implementation through the widely adopted QuPath software, we aim to greatly broaden access.

The WSInfer Python runtime is preferable for batch processing large numbers of slides, for example in a large-scale study. The results can be exported in a QuPath-compatible format for visualization. Direct use of the QuPath extension, however, means that it is also possible for a QuPath user to interactively select regions of interest and obtain results for any slide immediately, without leaving the software. We anticipate that making the application of models more streamlined in this way will encourage more pathologists to try the methods on new data. This should, in turn, make it easier to identify strengths and weaknesses, and thereby accelerate the critical feedback loop necessary to develop robust and generalizable algorithms.

Several tools exist for deploying trained models on whole slide images, including TIA Toolbox (11), MONAI (12), SlideFlow (13), and PHARAOH (14). WSInfer complements these by specifically targeting highly optimized, user-friendly support for patch based WSI inference methods. We expect that these tools may be used together and are keen to promote interoperability. To this end, the WSInfer Model Zoo implements a minimal model configuration specification that accompanies each trained model, with the intention that it may be used by other software beyond the direct WSInfer ecosystem. We host several trained patch classification models in the zoo, including two models from TIA Toolbox, and intend to incorporate more models in future work.

It is important to note that WSInfer itself supports a variety of patch classification models, but is agnostic to a user's choice of model. It is intended for research use only, and we make no claims regarding the suitability of the models for specific applications. Hence, users assume the responsibility of verifying the suitability of any model for their purposes. Indeed, it is our expectation that promising digital pathology methods will often be found not to perform well on new images; generalization across cohorts, scanners, and laboratories is a hard problem. However, we believe that an important first step to addressing this must be to enable existing models to be properly scrutinized by the research community, to identify what does and does not work. We hope that WSInfer may prove useful in this regard.

## Acknowledgements

The development of the WSinfer infrastructure by the Stony Brook authors was supported by Stony Brook Provost ProFund 2022 award and through the generosity of Bob Beals and Betsy Barton. JRK was also supported by the National Institutes of Health grant T32GM008444 (NIGMS) and by the Medical Scientist Training Program at Stony Brook University. The QuPath WSInfer extension was developed by the Edinburgh authors and was made possible in part by grant number 2021- 237595 from the Chan Zuckerberg Initiative DAF, an advised fund of Silicon Valley Community Foundation. This research was funded in part by the Wellcome Trust 223750/Z/21/Z. The results shown here are in whole or part based upon data generated by the TCGA Research Network: https://www.cancer.gov/tcga. For the purpose of open access, the author has applied a Creative Commons Attribution (CC BY) license to any Author Accepted Manuscript version arising from this submission.

# Figures

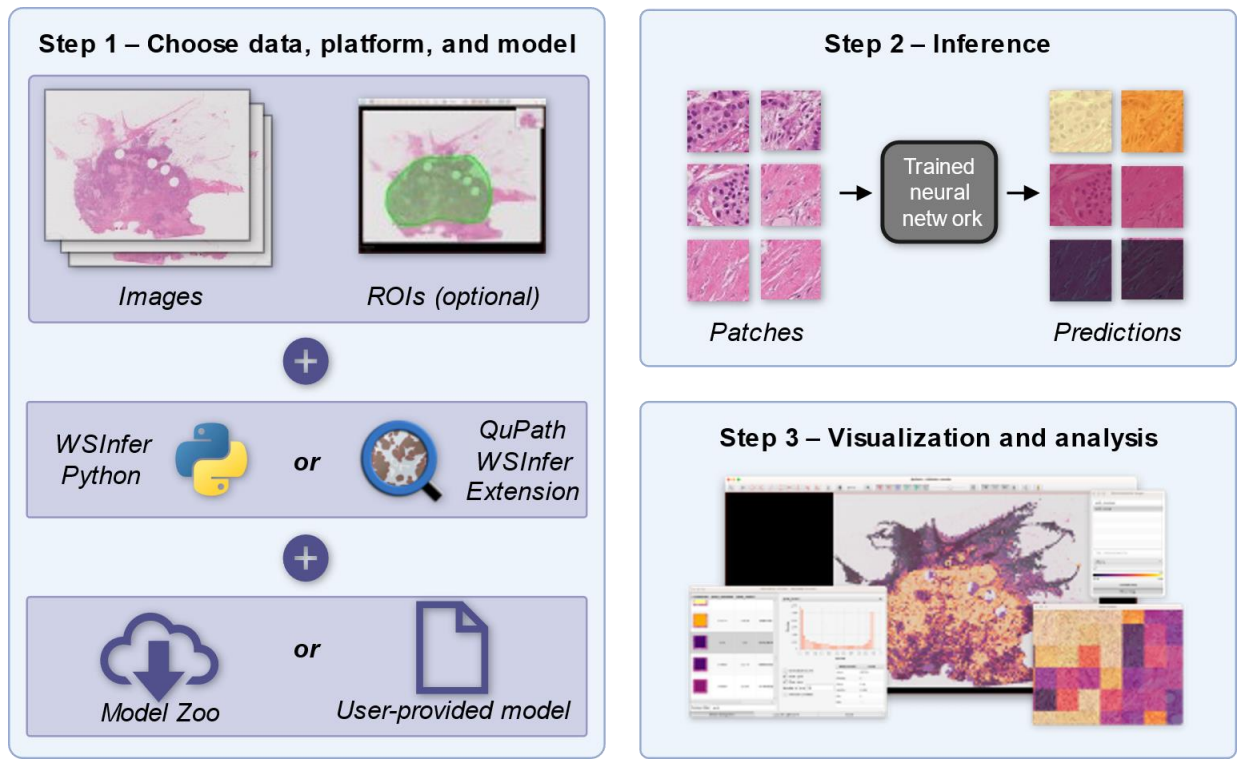

**Figure 1.** The WSInfer ecosystem streamlines the deployment of trained deep neural networks on whole slide images through three steps. In Step 1, users begin by selecting their WSIs and specifying the platform for model inference along with the choice of a pretrained model. If employing the WSInfer Python Runtime, the dataset is expected to be a directory containing WSI files. Alternatively, when using the WSInfer QuPath extension, the image currently open in QuPath serves as the input. QuPath users also have the option to designate a region of interest for model inference. The pretrained model can be selected from the WSInfer Model Zoo or users can provide their own model in TorchScript format. In Step 2, WSInfer performs a series of processing steps, including the computation of patch coordinates at the patch size and spacing prescribed by the model. Image patches are loaded directly from the WSI and used as input to the patch classification model. The model outputs are aggregated and saved to CSV and GeoJSON files. In Step 3, model outputs can be visualized and analyzed in QuPath or other software. This example shows breast tumor patch classification on a slide from TCGA.